\documentclass[pre,twocolumn,10pt,aps,longbibliography]{revtex4-1}

 \usepackage[utf8]{inputenc}
 \usepackage[greek,english]{babel}
 \usepackage{bm}
 \usepackage{verbatim}
 \usepackage{amsmath}
 \usepackage{amssymb}
 \usepackage{amsthm}
 \usepackage{latexsym}
 \usepackage{amsfonts}
 \usepackage{epsfig}
 \usepackage{epstopdf}
 \usepackage{wasysym} 
 \usepackage{subfigure}
 \usepackage{color}
 \definecolor{darkblue}{rgb}{0,0,.5}
 \definecolor{BLUE}{rgb}{0,0,1}
 \definecolor{BLACK}{rgb}{0,0,0}
 \usepackage[linktocpage, colorlinks=true, linkcolor=darkblue, citecolor=darkblue]{hyperref}
 \usepackage[all]{hypcap}
 \usepackage[makeroom]{cancel}

\newcommand{\bb}[1]{\textbf{#1}}

\begin{document}

\title{Comment on ``Extending the Laws of Thermodynamics for Arbitrary Autonomous Quantum Systems''}

\author{Philipp Strasberg}
\affiliation{F\'isica Te\`orica: Informaci\'o i Fen\`omens Qu\`antics, Departament de F\'isica, Universitat Aut\`onoma de Barcelona, 08193 Bellaterra (Barcelona), Spain}

\date{\today}

\begin{abstract}
 Recently, Elouard and Lombard Latune [PRX Quantum \bb{4}, 020309 (2023)] claimed to extend the laws of 
 thermodynamics to ``arbitrary quantum systems'' valid ``at any scale'' using ``consistent'' definitions 
 allowing them to ``recover known results'' from the literature. I show that their definitions are in conflict 
 with textbook thermodynamics and over- or underestimate the real entropy production by orders of magnitude. 
 The cause of this problem is traced back to problematic definitions of entropy and temperature, the latter, for 
 instance, violates the zeroth law. It is pointed out that another framework
 presented in PRX Quantum \bb{2}, 030202 (2021) does not suffer from these problems, while Elouard and Lombard Latune 
 falsely claim that it only provides a positive entropy production for a smaller class of initial states.
 A simple way to unify both approaches is also presented.
\end{abstract}

\maketitle

\newtheorem{lemma}{Lemma}[section]


A recent interesting attempt of Elouard and Lombard Latune (abbreviated ELL in the following) suggests microscopic 
definitions for thermodynamic quantities for two (``arbitrary'' and of ``any scale'') interacting quantum systems $A$ 
and $B$~\cite{ElouardLombardLatunePRXQ2023}. Their legitimate starting point is that the traditional dichotomy of heat 
and work reservoirs should be contained as limiting cases in a fully quantum description, and they succeed to derive 
formal mathematical identities resembling known relations from textbook thermodynamics. In the following, I will 
explain that this resemblance is, at best, only formal. 

For pedagogical purposes I start with the identity resembling Clausius' inequality for a system $A$ in contact with a 
single heat bath,
\begin{equation}\label{eq Clausius}
 \Delta S_A - \int_0^t dt' \beta_B(t')\dot Q_B(t') \ge 0.
\end{equation}
It is derived by ELL, see Eq.~(18) in Ref.~\cite{ElouardLombardLatunePRXQ2023}, for \emph{any} decorrelated initial 
state of the form $\rho_A(0)\otimes\rho_B(0)$ by identifying $S_A(t) \equiv S_\text{vN}[\rho_A(t)]$ (with $S_\text{vN}$ 
the von Neumann entropy), by defining the inverse temperature $\beta_B(t)$ of $B$ by equating 
$S_\text{vN}[\rho_B(t)] = S_\text{vN}[w_B(\beta_B(t))]$, where $w_B(\beta_B(t))$ denotes the Gibbs (canonical) state 
of $B$, and by setting $\dot Q_B(t) \equiv -\dot S_\text{vN}[\rho_B(t)]/\beta_B(t)$. 

Elementary algebra shows that these definitions make Eq.~(\ref{eq Clausius}) identical to 
\begin{equation}\label{eq MI}
 I_{AB}(t) \ge 0,
\end{equation}
where $I_{AB}(t) = S_\text{vN}[\rho_A(t)] + S_\text{vN}[\rho_B(t)] - S_\text{vN}[\rho_{AB}(t)]$ is the non-negative 
mutual information. This quantity is upper bounded by $2\min\{\ln d_A,\ln d_B\}$, where $d_A$ 
($d_B$) denotes the Hilbert space dimension of $A$ ($B$). Thus, whenever either $d_A$ or $d_B$ is small, 
the ELL entropy production in Eq.~(\ref{eq MI}) is bounded by a small number. 

Problematic results follow, which are best illustrated with extreme cases~\footnote{Note that ELL emphasize that their
framework is ``universally valid'' for ``arbitrary'' systems $A$ and $B$ and ``equivalent to earlier well-accepted
expressions''}. For instance, consider the case where $B$ is very small (say a single qubit) and $A$ describes a box of
volume $V$ with $N$ gas particles initially confined to a smaller volume $V'<V$. Then, the expansion of the gas generates
an entropy production proportional to $N$, whereas the ELL entropy production can never exceed $2\ln2$ (in units of $k_B$).
Similar problems appear if $A$ is very small (say again a single qubit) and $B$ is very large. For instance, let $A$ be
driven for a long time by a work reservoir (e.g., an external laser field or a part of $B$ that autonomously implements
a driving source). Then, entropy production should grow with time, but the ELL entropy production can again never exceed
$2\ln2$. Further counterexamples can be constructed by assuming some intrinsic dissipation in $B$, e.g., $B$ could be
composed of different regions initialized with different temperatures, something which is clearly conceivable for arbitrary
initial states $\rho_B(0)$, among other examples. Note that the importance of entropy production to scale extensively has
been emphasized in quantum thermodynamics~\cite{PtaszynskiEspositoPRL2019}.

The reason for this conflict can be traced back to the ELL definitions of entropy and temperature. Since the inadequacy 
of von Neumann entropy has been discussed at length in the literature, including recent 
references~\cite{SafranekDeutschAguirrePRA2019b, GoldsteinEtAlInBook2020, StrasbergWinterPRXQ2021}, I here only focus 
on the ELL definition of temperature, which---to the best of my knowledge---was first introduced in 
Ref.~\cite{AlickiFannesPRE2013}. Recall that the ELL temperature is determined by equating the von Neumann entropies 
of the actual state of $B$ with a corresponding Gibbs ensemble. It follows, first, that this definition requires 
\emph{full} knowledge of the state of $B$ (as also acknowledged by ELL). Second, ELL temperature is not 
always defined if the ground state of $B$ is degenerate, in contrast to the claim that it ``always admits a unique 
solution''~\cite{ElouardLombardLatunePRXQ2023}. Excluding this in the following, it follows that ELL temperature is 
\emph{always zero} if the state of $B$ is \emph{pure}---\emph{independent} of its energy, which is certainly
problematic if one wants to reproduce the equation $1/T = \partial_E S(E)$. It also conflicts with
empirically measured temperatures and it is unable to predict the flow of energy. To be specific, consider two 
\emph{equal} (in size and constituents) bodies in thermal contact. Then, the flow of energy is determined by
their energy difference, which is not predicted by the ELL temperature~\footnote{Note that ELL derive a relation 
showing that some \emph{redefined ELL heat} flows along the ELL temperature gradient, but this ELL heat does not equal 
the flow of energy between the bodies (not even in the long time limit). Indeed, one can show that the change in 
von Neumann entropy of system $B$ is upper bounded by $3\ln d_A$ if $d_A\le d_B$. This strong bound sets again 
problematic bounds for the change in ELL temperature and ELL heat.}. In addition, suppose that these bodies have
initially the same energy such that there is no net heat flow and their empirically measured temperature stays 
constant---in contrast, the ELL temperature would \emph{always change} due to the build up of
correlations between $A$ and $B$. 

Coming back to entropy production, I remark that in most parts ELL discuss a limiting case of Eq.~(\ref{eq Clausius}) 
obtained by replacing $\beta_B(t)$ by $\beta_B(0)$ for all $t$, see Eq.~(5) in 
Ref.~\cite{ElouardLombardLatunePRXQ2023}. Without going into details, I only remark that Eq.~(5) in
Ref.~\cite{ElouardLombardLatunePRXQ2023} also suffers from serious drawbacks. It suffices to point out, for instance, that
it \emph{diverges} whenever system $B$ is initially in a pure state.

Instead, I find it important to make another point. In phenomenological thermodynamics the most general formulation 
of the second law stipulates that thermodynamic entropy in an isolated system does not decrease. Clausius' inequality 
in Eq.~(\ref{eq Clausius}) is a \emph{specific application} of this second law, meant to apply only when a thermal 
bath can be identified whose change in thermodynamic entropy equals heat divided by temperature or, more operationally 
speaking, if the energy of the bath is the only accessible quantity. Therefore, for a completely general and fully known
system $B$ as considered by ELL one would not even expect that the second law can and should be expressed in form of
Eq.~(\ref{eq Clausius}); compare also with Ref.~\cite{JaynesInBook1992} for a discussion on how the form of the 
second law changes depending on ones state of knowledge. 

A framework that does not suffer from the drawbacks mentioned above 
exists~\cite{StrasbergWinterPRXQ2021, RieraCampenySanperaStrasbergPRXQ2021, StrasbergDiazRieraCampenyPRE2021, 
StrasbergBook2022, StokesPRE2022}. ELL complain that the expression for entropy production in this framework is 
positive only for a smaller class of initial states, but that is an unfortunate misrepresentation of this framework. 
As explicitly discussed, e.g., in Ref.~\cite{StrasbergWinterPRXQ2021}, if one assumes perfect knowledge 
of a system (as done by ELL), it becomes legitimate to use von Neumann entropy as thermodynamic entropy. In this case, 
the second law of Ref.~\cite{StrasbergWinterPRXQ2021} reduces to 
\begin{equation}
 \Delta S_A + \Delta S_B \ge 0,
\end{equation}
which is numerically \emph{identical} to the expression provided by ELL in Eq.~(\ref{eq MI}). It therefore also holds 
for the \emph{same} class of initial states. However, for the reasons clearly spelled out in the previous paragraph, 
Ref.~\cite{StrasbergWinterPRXQ2021} refrains from identifying $\Delta S_B$ with some heat flow divided by temperature, 
whereas ELL try to press their second law into a form that is inadequate from a thermodynamic point of view. 

Finally, I stress the well known fact that the second law is of \emph{statistical nature}. Superiority claims based on
the mathematical truth that some inequality \emph{strictly} holds for a larger class of initial states might not reflect 
any physical truth. Indeed, by time reversal symmetry violations of Eq.~(\ref{eq Clausius}) \emph{must be possible}, 
and all what one can hope for is that an increase in Eq.~(\ref{eq Clausius}) is much more likely than a decrease for 
arbitrary states (a result which is not rigorously proven but can be anticipated from Refs.~\cite{VonNeumann1929, 
VonNeumannEPJH2010}). Already this basic insight casts doubts on the validity of the ELL framework, which predicts that 
a spontaneous descrease in Clausius' expression is impossible (and not just extremely unlikely) for all 
decorrelated states. 

To conclude, by using rigorous identities and simple relevant examples, I have shown that the ELL framework contradicts
much what we know about statistical mechanics and thermodynamics. Importantly, the points I have made have not been
noticed in Ref.~\cite{ElouardLombardLatunePRXQ2023}. How far the ELL framework is useful for small systems, whose states
are perfectly known, remains to be established. Nonetheless, the general question addressed by ELL deserves further
attention, some vague steps towards a more realistic description have been presented by ELL themselves in Appendix
D~\cite{ElouardLombardLatunePRXQ2023}. Personally, and in unison with Ref.~\cite{JaynesInBook1992}, I believe it is 
better to approach this question based on a rigorous and flexible definition of entropy, which is consistent with what we 
know already. The first steps to do so have been layed down in Refs.~\cite{SafranekDeutschAguirrePRA2019b, 
StrasbergWinterPRXQ2021, RieraCampenySanperaStrasbergPRXQ2021, StrasbergDiazRieraCampenyPRE2021, StrasbergBook2022, 
StokesPRE2022, VonNeumann1929, VonNeumannEPJH2010}. Indeed, it is straightforward to adapt the temperature definition
of Refs.~\cite{ElouardLombardLatunePRXQ2023, AlickiFannesPRE2013} to the observational entropy used in
Refs.~\cite{SafranekDeutschAguirrePRA2019b, StrasbergWinterPRXQ2021, RieraCampenySanperaStrasbergPRXQ2021,
StrasbergDiazRieraCampenyPRE2021, StrasbergBook2022, StokesPRE2022, VonNeumann1929, VonNeumannEPJH2010}.
This would reduce to the framework of ELL for perfectly controlled quantum systems \emph{and} it would avoid the
criticism above if applied to a realistic many-body system. Whether this entropy-based definition of temperature is
better than its energy-based counterpart~\cite{StrasbergWinterPRXQ2021, StrasbergDiazRieraCampenyPRE2021,
RieraCampenySanperaStrasbergPRXQ2021, StrasbergBook2022} is a subtle question worthwhile to pursue.

\subsection*{Acknowledgements}

PS is financially supported by ``la Caixa'' Foundation (ID 100010434, fellowship code LCF/BQ/PR21/11840014)
and acknowledges further support from the European Commission QuantERA grant ExTRaQT (Spanish MICINN project
PCI2022-132965), by the Spanish MINECO (project PID2019-107609GB-I00) with the support of FEDER funds, the Generalitat
de Catalunya (project 2017-SGR-1127) and by the Spanish MCIN with funding from European Union NextGenerationEU
(PRTR-C17.I1).



\bibliography{/home/philipp/Documents/references/books,/home/philipp/Documents/references/open_systems,/home/philipp/Documents/references/thermo,/home/philipp/Documents/references/info_thermo,/home/philipp/Documents/references/general_QM,/home/philipp/Documents/references/math_phys,/home/philipp/Documents/references/equilibration,/home/philipp/Documents/references/time,/home/philipp/Documents/references/cosmology,/home/philipp/Documents/references/general_refs}

\end{document}